\pdfoutput=1
\documentclass{article}
\usepackage[sort,authoryear]{natbib}
\usepackage{mathtools}
\usepackage{amssymb}
\usepackage{amsthm}
\usepackage{enumitem}
\usepackage[margin=1in]{geometry}
\usepackage{graphicx}
\usepackage{subcaption}
\usepackage[utf8]{inputenc}
\usepackage[english]{babel}
\usepackage{xcolor}
\usepackage[colorlinks = true, citecolor = blue, urlcolor = blue]{hyperref}
\PassOptionsToPackage{hyphens}{url}\usepackage{hyperref}
\usepackage{indentfirst}
\usepackage{apptools}
\usepackage{cleveref}
\usepackage{algorithm}
\usepackage[noend]{algpseudocode}

\graphicspath{{fig/}}

\AtAppendix{\counterwithin{lemma}{section}}
\AtAppendix{\counterwithin{proposition}{section}}

\DeclareMathOperator*{\argmax}{argmax}

\newtheorem{theorem}{Theorem}
\newtheorem{corollary}{Corollary}

\newcommand{\Cov}{\textup{Cov}}
\newcommand{\Var}{\textup{Var}}
\newcommand{\R}{\mathbb{R}}

\title{Powerful rank verification for multivariate Gaussian data with any covariance structure}
\author{Anav Sood \\ Stanford University}
\date{March 1st, 2025}

\begin{document}

\maketitle

\begin{abstract}

Upon observing $n$-dimensional multivariate Gaussian data, when can we infer that the largest $K$ observations came from the largest $K$ means? When $K=1$ and the covariance is isotropic, \cite{Gutmann} argue that this inference is justified when the two-sided difference-of-means test comparing the largest and second largest observation rejects. Leveraging tools from selective inference, we provide a generalization of their procedure that applies for both any $K$ and any covariance structure. We show that our procedure draws the desired inference whenever the two-sided difference-of-means test comparing the pair of observations inside and outside the top $K$ with the smallest standardized difference rejects, and sometimes even when this test fails to reject. Using this insight, we argue that our procedure renders existing simultaneous inference approaches inadmissible when $n > 2$. When the observations are independent (with possibly unequal variances) or equicorrelated, our procedure corresponds exactly to running the two-sided difference-of-means test comparing the pair of observations inside and outside the top $K$ with the smallest standardized difference.
\end{abstract}

\section{Introduction and results}
\label{sec:intro}

\subsection{Motivation}

Having observed data, it is often natural to ask whether the best observation actually came from the best population. As motivation, we present two important real world variants of this problem. \newline 

\noindent \textbf{Rank verification for large language models: } Chatbot Arena \citep{Chiang} is a platform that currently has over thirty-thousand daily users and ranks the performance of $n=206$ large language models according to user preference data. Is the model at the top of the leaderboard actually the best model?  \newline 

\noindent \textbf{Rank verification in multi-arm clinical trials: } In multi-arm clinical trials, each patient is randomly assigned to receive one of $n$ different treatments (including a control). Is the treatment with the largest observed average treatment effect actually the best treatment? \newline 

Both of these scenarios are examples of a rank verification problem, and they can be formalized as follows. We observe a multivariate Gaussian vector $X \sim N(\mu, \Sigma)$ and see that $W = \argmax_{1 \leq i \leq n} X_i$ is the index of the largest observation. Can we claim that $\mu_{W} > \max_{j \neq W} \mu_j$, i.e., that the largest observation came from the largest mean? We elaborate on how each of the above scenarios reduces to solving this problem.\newline 

\noindent \textbf{Rank verification for large language models: } The Chatbot Arena dataset is constructed by asking users which of two models performed better on a prompt. Treating these responses as i.i.d samples, the leaderboard fits a Bradley-Terry model \citep{Bradley} and uses the resulting coefficients $\hat{\beta} \in \R^n$ to rank the models. These coefficients follow a central limit theorem when the number of samples is large, i.e., $\hat{\beta} \dot \sim N(\beta, \Sigma)$ for some non-diagonal covariance $\Sigma$. Letting $W$ be the index of the model with the largest fitted coefficient, we want to know, is $\beta_{W} > \min_{j \neq W} \beta_j$? \newline 

\noindent \textbf{Rank verification in multi-arm clinical trials: } If there are enough participants in the clinical trial, then the the treatments' average observed effects $X \in \R^n$ obey a central limit theorem, i.e., $X \dot \sim N(\mu, \Sigma)$ with diagonal covariance $\Sigma$. We want to know, is $\mu_{W} > \max_{j \neq W} \mu_j$? \newline 

Briefly, we mention that our formalization also encompasses the problem of verifying that the machine learning model with the best performance on a challenge dataset is actually the best model. In this case, the $n$ models' observed average performances $X \in \R^n$ on the dataset obey a central limit theorem (provided that the challenge dataset is moderately large), i.e., $X \sim N(\mu, \Sigma)$. Because the same dataset is used to evaluate the models, the $X_i$ are correlated and $\Sigma$ may be highly non-diagonal. Again we want to know, is $\mu_{W} > \max_{j \neq W} \mu_j$?. 

This paper considers a generalization of our motivating problem. Defining $S$ to be the set containing the indices of the largest $K < n$ entries of $X$, we aim to draw the inference $\min_{i \in S} \mu_i - \max_{j \not \in S} \mu_j > \delta$ that the largest $K$ observations came from means that are more than $\delta$ larger than the rest. We recover our motivating problem by setting $K=1$ and $\delta=0$. 

The main contribution of this paper is to provide a powerful and computationally tractable error controlling procedure for drawing the inference $\min_{i \in S} \mu_i - \max_{j \not \in S} \mu_j > \delta$. For a pre-specified level $\alpha$, the probability of our procedure falsely drawing this inference will be at most $\alpha$. Throughout our discussion, we will assume that the covariance $\Sigma$ is known. In practice, $\Sigma$ is not known but can be estimated from the data. 

\subsection{Method and theoretical results}

In this subsection, we summarize the results of the paper. We imagine observing $n$-dimensional data $X \sim N(\mu, \Sigma)$ where $\Sigma$ is known. Our only restriction on $\Sigma$ is that we require $X_i$ and $X_j$ to be not perfectly correlated when $i \neq j$. Considering the null hypotheses $H^{\delta}_{ij} : \mu_i - \mu_j \leq \delta$, we derive an error controlling procedure for rejecting the data dependent union null $\cup_{i \in S, j \not \in S}  H^{\delta}_{ij}$, where $S$ is the set of the largest $K$ observations' indices. Formally, a false rejection happens when we reject $\cup_{i \in S, j \not \in S}  H^{\delta}_{ij}$ and $\mu_i$ is not more than $\delta$ larger than $\mu_j$ for all $i \in S$ and $j \not \in S$. Tying back to our motivation, we can safely draw the inference $\min_{i \in S} \mu_i - \max_{j \not \in S} \mu_j > \delta$ when we reject $\cup_{i \in S, j \not \in S}  H^{\delta}_{ij}$.

To simplify our exposition, we present our results in the special case that $\delta=0$, i.e., we just consider the problem of rejecting $\cup_{i \in S, j \not \in S}  H^{0}_{ij}$ and verifying $\min_{i \in S} \mu_i > \max_{j \not \in S} \mu_j$. The most general versions of our results are clearly stated in \Cref{sec:proofs}, where we provide proofs of our claims as well.

Prior to stating our simplified results, we introduce some notation. First, for a pair $i \neq j$, we define 
\begin{equation*}
    D_{ij} = \frac{X_i - X_j}{v_{ij}}, \quad v^2_{ij} = \Var(X_i - X_j) = \Sigma_{ii} - 2\Sigma_{ij} + \Sigma_{jj}
\end{equation*}
to be the standardized difference between $X_i$ and $X_j$. Considering another pair $k \neq \ell$, we use $\rho_{ij, k\ell}$ to denote the correlation between $D_{ij}$ and $D_{k\ell}$.

Using this notation, \Cref{thm:method} states our method. Though it may look complicated, it is easy to implement on a computer and we will soon see that its behavior is very interpretable. In our statement of the theorem, we adopt the convention that the minimum of an empty set is $\infty$. We provide a proof of a generalized version of \Cref{thm:method} that applies for any $\delta \in \R$ in \Cref{sec:method_proof}. 

\begin{theorem}[Gaussian rank verification]
    \label{thm:method}
    If we reject $\cup_{i \in S, j \not \in S}  H^0_{ij}$ when 
    \begin{equation}
        \label{eq:method}
        \max_{i \in S, j \not \in S} \frac{ \left[1 - \Phi(D_{ij}) \right] - \left[1 - \Phi\left(\min\limits_{\substack{k \in S, \ell \not \in S: \\ \rho_{ij, k\ell} < 0}} D_{ij} - \frac{1}{\rho_{ij, k\ell}} D_{k \ell} \right) \right] }{\left[1 - \Phi\left(\max\limits_{\substack{k \in S,\, \ell \not \in S: \\ \rho_{ij, k\ell} > 0}} D_{ij} - \frac{1}{\rho_{ij, k\ell}} D_{k \ell} \right) \right] -  \left[1 - \Phi\left(\min\limits_{\substack{k \in S, \ell \not \in S: \\ \rho_{ij, k\ell} < 0}} D_{ij} - \frac{1}{\rho_{ij, k\ell}} D_{k \ell} \right) \right]}\leq \alpha,
    \end{equation}
    then, conditional on $S$, the probability of making a false rejection is at most $\alpha$. 
\end{theorem}

Our next result, \Cref{thm:interpretation}, helps us make sense of \Cref{thm:method}'s method. We prove an analog of \Cref{thm:interpretation}
that applies whenever $\delta \geq 0$ in \Cref{sec:interpretation_proof}. 

\begin{theorem}[Understanding Gaussian rank verification]
    \label{thm:interpretation} Let 
     $I \in S$ and $J \not \in S$ be the indices of any pair of observations inside and outside of the top $K$ that achieve the smallest possible standardized difference, i.e., $D_{IJ} = \min_{i \in S, j \not \in S} D_{ij}$. Then the procedure from \Cref{thm:method} is guaranteed to reject $\cup_{i \in S, j \not \in S}  H^{0}_{ij}$ whenever 
    \begin{equation}
        \label{eq:interpretation}
        1 - \Phi(D_{IJ}) \leq \alpha/2.
    \end{equation}
\end{theorem} 

Essentially, \Cref{thm:interpretation} tells us that \Cref{thm:method}'s test will reject $\cup_{i \in S, j \not \in S}  H^0_{ij}$ whenever the two-sided difference-of-means test comparing the pair of observations inside and outside of the top $K$ with the smallest standardized difference rejects, and possibly also in other situations as well. We mention that the time complexity of running \Cref{thm:method}'s test is $O(K^2(n-K)^2)$, which can be $O(n^4)$ in the worst case (e.g. if $K = n/2$). This will still not be prohibitive for many problems, but if it is, \Cref{thm:interpretation} tells us that we could also safely reject $\cup_{i \in S, j \not \in S}  H^0_{ij}$ whenever $1 - \Phi(D_{IJ}) \leq \alpha/2$, a condition that only takes $O(K(n-K))$ time to check (which is $O(n^2)$ in the worst case). Because \Cref{thm:method}'s test can reject even when this condition does not hold, however, doing so can result in a loss of power. In adversarially chosen settings, this loss of power can be very large (see \Cref{sec:counterexample_appdx} for a numerical example). We mention that, when $K=1$, as in our original motivating problem, \Cref{thm:method}'s test only takes $O(n^2)$ time to run and \Cref{thm:interpretation}'s condition only takes $O(n)$ time to check.

\Cref{thm:interpretation} also clarifies a sense in which \Cref{thm:method}'s error control is tight. Fixing some covariance $\Sigma$, if we set $\infty = \mu_1 = \dots \mu_{K-1} > \mu_{K} = \mu_{K + 1} > \mu_{K+2} = \dots = \mu_n= -\infty $ , then $\cup_{i \in S, j \not \in S}  H^0_{ij}$ is always true and any rejection is false. In this example, it will always be the case that $I = K$, $J= K+ 1$, and the two-sided difference-of-means test comparing 
$X_K$ to $X_{K+1}$ will falsely reject with probability exactly $\alpha$ (after all, this is the setting of a vanilla two-sided test). \Cref{thm:method}'s test both (1) rejects whenever this difference-of-means test does and (2) still maintains error control, so it must falsely reject with probability exactly $\alpha$ as well. For the generalized version of \Cref{thm:method} that works for any $\delta \in \R$, the same tightness can be achieved by setting $\mu_K$ and $\mu_{K+1}$ to be exactly $\delta$ apart.

\Cref{cor:reduction} tells us that, when the data is independent or equicorrelated, \Cref{thm:method}'s test rejects $\cup_{i \in S, j \not \in S} H^0_{ij}$ \underline{exactly} when \Cref{thm:interpretation}'s condition \eqref{eq:interpretation} is satisfied, i.e., rejecting $\cup_{i \in S, j \not \in S} H^0_{ij}$ according to \Cref{thm:interpretation}'s condition results in no power loss. Also, in \Cref{sec:special_appdx}, we show for the equicorrelated case that the indices $I$ and $J$ from the condition \eqref{eq:interpretation} are always those of the $K$ and $(K+1)$st largest observations. We provide a proof of \Cref{cor:reduction} in \Cref{sec:reduction_proof}. It is specific to the special case $\delta=0$, and has no analog when $\delta \neq 0$.  

\begin{corollary}[Exact equivalence]
    \label{cor:reduction}
    If $\Sigma$ is diagonal or an equicorrelation matrix (i.e., $\Sigma_{ij}$ is $\rho \sigma^2$ when $i = j$ and $\rho$ when $i \neq j$), then the procedure from \Cref{thm:method} rejects $\cup_{i \in S, j \not \in S}  H^{0}_{ij}$ if and only if the condition \eqref{eq:interpretation} from \Cref{thm:interpretation} is satisfied. 
\end{corollary}

There are a couple other notable situations where the conclusion of \Cref{cor:reduction} applies. When $K=1$ or $K=n-1$ and the $X_i$ have a small amount of autocorrelation (i.e., $\Sigma_{ij} = \sigma^2\rho^{|i-j|}$ with $|\rho| \leq 1/2)$, the result of \Cref{cor:reduction} still holds. It also holds when $K=1$ and we use a multivariate Gaussian distribution to approximate the joint distribution of $t$ multinomial trials $Y \sim \text{Multinomial}(t, \pi_1, \dots, \pi_n)$, i.e., we define $\hat{\pi} = Y/t$ and apply our method to $\hat{\pi} \dot \sim N(\pi, \hat{\pi}/t - \hat{\pi} \hat{\pi}^{\top}/t)$. \Cref{sec:special_appdx} discusses these settings in more detail. 

Now that we have a good grasp of \Cref{thm:method}'s test and its behavior, we can establish why it is a surprisingly powerful procedure. For our rank verification problem, the natural alternative to our selective approach is to perform simultaneous inference. The most standard example of this is Tukey's honestly significant difference (HSD) test \citep{Tukey}. Using the random indices $I$ and $J$ from \Cref{eq:interpretation}, Tukey's test, once adapted to our problem, would tell us to reject $\cup_{i \in S, j \not \in S} H^0_{ij}$ whenever
\begin{equation*}
    X_I \geq X_J + v_{IJ} h_{1-\alpha}, \qquad h_{1-\alpha} = \text{Quantile}\left(1-\alpha, \max_{i \neq j} \frac{|Z_i - Z_j|}{v_{ij}} \right) \text{ with } Z = X - \mu.
\end{equation*}
In contrast, \Cref{thm:method}'s test is guaranteed to reject $\cup_{i \in S, j \not \in S} H^0_{ij}$ whenever 
\begin{equation*}
    X_I \geq X_J + v_{IJ} z_{1-\alpha/2}, \qquad z_{1-\alpha} = \text{Quantile} \left(1-\alpha, Z \right) \text{ with } Z \sim N(0, 1)
\end{equation*}
When $n=2$, these two approaches coincide. But as soon as $n > 2$, the quantile $h_{1-\alpha}$ becomes strictly larger than $z_{1-\alpha/2}$, and our test's rejection region becomes a strict superset of Tukey's HSD rejection region. In the case that the $X_i$ are independent, the growth of $h_{1-\alpha}$ is at least on the order of $\sqrt{\log n}$ (see \Cref{sec:simul_appdx} for justification). The quantile $z_{1-\alpha/2}$ that our procedure uses, however, stays fixed. In essence, our approach avoids a multiple comparisons correction that cannot be avoided by simultaneous inference

Drawing from the prior rank verification literature \citep{Bofinger1983,Bofinger1985,Hsu1981,Hsu1984}, there is a variant of Tukey's HSD that is more powerful for our specific rank verification problem (although, to the best of our knowledge, it is not computationally tractable when $\Sigma$ is not isotropic). We provide an analogous discussion for this variant in \Cref{sec:simul_appdx}. The story remains is identical. When $n > 2$ our procedure avoids a multiple comparisons correction that the this more powerful simultaneous procedure cannot, and our procedure's rejection region remains a strict superset of even this more powerful simultaneous approach's rejection region. 

The remainder of the article is devoted to proving more general versions of the results in this section. One of them is a generalization of \Cref{thm:method}'s method that applies for any $\delta \in \R$, not just $\delta =0$. We argue in \Cref{sec:clb_appdx} that the smallest $\delta$ for which this generalized method fails to reject $\cup_{i \in S, j\not \in S} H^{\delta}_{ij}$ provides a $1-\alpha$ confidence lower bound for the gap $\min_{i \in S} \mu_i - \max_{j \neq S} \mu_j $ between the smallest mean in the selected set and the largest mean in the unselected set that is valid conditional on $S$. In practice, this $\delta$ can be found via a binary search. \Cref{sec:clb_appdx} also discusses how to leverage a more general version of \Cref{thm:interpretation} to get a less powerful, but more computationally easier confidence lower bound for this quantity.

\subsection{Related work}

\cite{Gutmann} study our problem in the case that $K=1$, $\delta=0$, and the data $X_i \sim N(\mu_i, \sigma^2)$ are independent Gaussian samples with common variance. They show that drawing the inference $\min_{i \in S} \mu_i - \max_{j \not \in S} \mu_j > \delta$ whenever the two-sided difference-of-means test comparing the largest and second largest observation rejects is an error controlling procedure. Our work provides a complete generalization of their result in the case of multivariate Gaussian data, allowing for any $K$, any covariance structure, and any $\delta \in \R$. Work prior to \cite{Gutmann} studied related rank verification problems in similar settings \citep{Bechhofer,Bofinger1983,Bofinger1985, Fabian,Hsu1981, Hsu1984,Gupta1956, Gupta1965, Desu1970}, but used simultaneous inference techniques and failed to avoid a multiplicity correction as \cite{Gutmann} did. Follow-up work to \cite{Gutmann} includes methods that avoid multiplicity corrections for other rank verification problems \citep{Maymin,Gutmann1987}, but in similarly restricted settings. Also, \cite{Cheng} extend \cite{Gutmann}'s result to the case of independent samples from a scale family with a monotone likelihood ratio (\cite{Gutmann} themselves handle the case of independent samples from a location family with a monotone likelihood ratio). 

To prove their result \cite{Gutmann}, condition on the index of the winning observation. This is a similar strategy to that of modern post-selection inference, a field initiated by the seminal work \cite{Lee}. By leveraging modern selective techniques, \cite{Hung} generalize \cite{Gutmann}'s procedure to apply for exponential families with Schur concave carrier measures (for $K=1$ and any $\delta \in \R$). Schur concavity requires the carrier measure to be symmetric, so for the multivariate Gaussian case \cite{Hung} only generalize \cite{Gutmann}'s procedure from the independent to the equicorrelated setting (our \Cref{cor:reduction} subsumes both cases). The main focus of \cite{Hung} is rank verification for multionmial data. If there are enough samples, then multinomial data can be approximated as correlated multivariate Gaussian data via the central limit theorem, and we show that our method behaves the same as theirs in \Cref{sec:special_appdx}. \cite{Hung} also consider rank verification for the Bradley-Terry model, but their method (1) only scales to games with roughly $n=40$ players and (2) requires each player to play the other players the same number of times (both conditions are violated in our Chatbot Arena motivating example). Work that is concurrent with and independent from ours considers multivariate Gaussian rank verification in the independent and unequal variance case \citep{Goldwasser}. Indeed, once restricted to this case, \Cref{thm:method}'s method matches that of \cite{Goldwasser}. \cite{Goldwasser}, however, do not show that the resulting method amounts to running the two-sided difference-of-means test comparing the observation inside the top $K$ and observation outside the top $K$ with the smallest standardized difference (i.e., they have no analog of \Cref{thm:interpretation} or \Cref{cor:reduction}). As a consequence, they do not formally characterize the method’s behavior or power. 

\section{Proofs}
\label{sec:proofs}

In this section, we prove more general versions of the results stated in \Cref{sec:intro}. The more general result we are aiming to prove is stated clearly at the start of each proof. We will use \begin{equation*}
    D^{\delta}_{ij} = \frac{(X_i - X_j) - \delta}{v_{ij}}
\end{equation*}
to denote the standardized distance between $X_i - X_j$ and $\delta$. Note that $D^0_{ij} = D_{ij}$, where $D_{ij}$ is the standardized difference between $X_i$ and $X_j$ from the previous section. 
 
\subsection{Proof of \Cref{thm:method}}
\label{sec:method_proof}

Considering the null hypotheses $H^{\delta}_{ij}: \mu_i - \mu_j \leq \delta $ we will show that rejecting the the data dependent union null $\cup_{i \in S, j \not \in S}  H^{\delta}_{ij}$ when
\begin{equation}
    \label{eq:rejection}
    \max_{i \in S, j \not \in S} \frac{ \left[1 - \Phi(D^{\delta}_{ij}) \right] - \left[1 - \Phi\left(\min\limits_{\substack{k \in S,\, \ell \not \in S: \\ \rho_{ij, k\ell} < 0}} D^{\delta}_{ij} - \frac{1}{\rho_{ij, k\ell}} D^0_{k \ell}\right) \right] }{\left[1 - \Phi\left(\max\limits_{\substack{k \in S,\, \ell \not \in S: \\ \rho_{ij, k\ell} > 0}} D^{\delta}_{ij} - \frac{1}{\rho_{ij, k\ell}} D^0_{k \ell} \right) \right] -  \left[1 - \Phi\left(\min\limits_{\substack{k \in S,\, \ell \not \in S \\ \rho_{ij, k\ell} < 0}} D_{ij}^{\delta} - \frac{1}{\rho_{ij, k\ell}} D_{k \ell}^0 \right) \right]}\leq \alpha.
\end{equation}
ensures that, conditional on $S$, the probability of a false rejection is at most $\alpha$.

Without loss of generality, we perform our analysis conditional on the specific event $S = \{1, \dots, K\}$ that the $X_1, \dots, X_K$ are larger than the $X_{K + 1}, \dots X_n$. Our strategy will mimic that presented in \cite{Hung} and \cite{Sood}. First, for pairs $i \leq K$ and $j > K$, we come up with a test for rejecting $H^{\delta}_{ij}$ that maintains error control conditional on $S = \{1, \dots, K\}$. We reject $\cup_{i \leq K, j > K}  H^{\delta}_{ij}$ when these tests reject for all $i \leq K$, $j > K$. Lemma 4 of \cite{Hung}, which is adopted from \cite{Berger}, ensures that in doing so we maintain error control conditional on $S = \{1, \dots, K\}$. 

Fix some $i \leq K$ and $j > K$. To design a test for rejecting $H^{\delta}_{ij}$ that maintains error control conditional on $S = \{1, \dots, K\}$, we will use the selective dominance machinery from \cite{Sood}. Normally, to maintain marginal Type I error control, we reject the null $H^{\delta}_{ij}$ using the p-value 
\begin{equation}
    \label{eq:p_value}
    p^{\delta}_{ij} = 1 - \Phi(D^{\delta}_{ij}).
\end{equation}
Defining
\begin{equation*}
    \epsilon^{\delta}_{ij, k \ell} = D^{0}_{k \ell} - \rho_{ij, k \ell} D^{\delta}_{ij}
\end{equation*}
it is straightforward to verify that the random vector $\epsilon^{\delta}_{ij}$, which consists of the entries $\epsilon^{\delta}_{ij, k \ell}$ for pairs $k\leq K$ and $\ell > K$, is independent of $D^{\delta}_{ij}$ and therefore also $p^{\delta}_{ij}$. Since the p-value \eqref{eq:p_value} corresponds to running a one-sided uniformly most powerful (UMP) test in a monotone likelihood ratio family (MLR), Example 3 in \cite{Sood} tells us that it is selectively dominant (see \cite[Definition 1]{Sood}) given $\epsilon^{\delta}_{ij}$. 

All that remains to do is characterize when we are selecting the p-value $p^{\delta}_{ij}$ to use for inference (i.e., when $S=\{1,\dots, K\}$, what values is this p-value taking?). Theorem 1 from \cite{Sood} then tells how to adjust the p-value to get a selective p-value. Rejecting $H^{\delta}_{ij}$ when this selective p-value is below $\alpha$ maintains error control conditional on  $S=\{1,\dots, K\}$. 

To do so, we consider $k\leq K$ and $\ell > K$ and rewrite the event 
\begin{align*}
    X_k > X_{\ell} &\iff D^{0}_{k \ell} > 0\\
              &\iff \epsilon^{\delta}_{ij, k\ell} + \rho_{ij, k\ell} D^{\delta}_{ij} > 0
\end{align*}
This leads to three cases:
\begin{enumerate}
    \item If $\rho_{ij, k \ell} > 0$ then $X_k > X_{\ell} \iff D_{ij}^{\delta} >  - \frac{1}{\rho_{ij, k \ell}} \epsilon^{\delta}_{ij, k\ell} $,
    \item If $\rho_{ij, k\ell} = 0$ then $X_k > X_{\ell} \iff \epsilon^{\delta}_{ij, k \ell} > 0$,
    \item If $\rho_{ij, k\ell} < 0$ then $X_k > X_{\ell} \iff D^{\delta}_{ij} <  - \frac{1}{\rho_{ij, k \ell}} \epsilon^{\delta}_{ij, k\ell}$
\end{enumerate}
Ultimately, we see that 
\begin{align*}
    S = \{1, \dots, K \} &\iff  X_k > X_{\ell} \text{ for all } k \leq K \text{ and } \ell > K  \\
    &\iff D^{\delta}_{ij} \in \left[\max_{\substack{k \leq K,\, \ell > K: \\ \rho_{ij, k\ell} > 0}} - \frac{1}{\rho_{ij, k\ell}}\epsilon^{\delta}_{ij, jk}, 
    \min_{\substack{k \leq K,\, \ell > K: \\ \rho_{ij, k\ell} < 0}} - \frac{1}{\rho_{ij, k\ell}}\epsilon^{\delta}_{ij, jk} \right] \text{ and } \min_{\substack{k \leq K,\, \ell > K: \\ \rho_{ij, k\ell} = 0}}\epsilon^{\delta}_{ij, k \ell} > 0.
\end{align*}
Essentially, the selection event $S=\{1, \dots, K\}$ corresponds to selecting $p^{\delta}_{ij}$ \eqref{eq:p_value} to use for inference when it lives in some closed interval $[A, B]$ with bounds that are a measurable function of $\epsilon^{\delta}_{ij}$. In this case, Theorem 1 from \cite{Sood} tells us that the selective p-value is $(p^{\delta}_{ij} - A)/(B-A)$. Writing everything out explicitly, the selective p-value is 
\begin{equation*}
    p^{\delta}_{sel, ij} = \frac{ \left[1 - \Phi(D^{\delta}_{ij}) \right] - \left[1 - \Phi\left(\min\limits_{\substack{k \leq K,\, \ell > K: \\ \rho_{ij, k\ell} < 0}} - \frac{1}{\rho_{ij, k\ell}}\epsilon^{\delta}_{ij, jk}\right) \right] }{\left[1 - \Phi\left(\max\limits_{\substack{k \leq K,\, \ell > K: \\ \rho_{ij, k\ell} > 0}} - \frac{1}{\rho_{ij, k\ell}}\epsilon^{\delta}_{ij, jk} \right) \right] -  \left[1 - \Phi\left(\min\limits_{\substack{k \leq K,\, \ell > K: \\ \rho_{ij, k\ell} < 0}} - \frac{1}{\rho_{ij, k\ell}}\epsilon^{\delta}_{ij, jk} \right) \right]}. 
\end{equation*}
Recalling the definition of $\epsilon_{ij, k\ell}$, we can rewrite this selective p-value as
\begin{equation}
    \label{eq:selective_p_value}
    p^{\delta}_{sel, ij} = \frac{ \left[1 - \Phi(D^{\delta}_{ij} ) \right] - \left[1 - \Phi\left(\min\limits_{\substack{k \leq K,\, \ell > K: \\ \rho_{ij, k\ell} < 0}} D^{\delta}_{ij} - \frac{1}{\rho_{ij, k\ell}} D^{0}_{k \ell} \right) \right] }{\left[1 - \Phi\left(\max\limits_{\substack{k \leq K,\, \ell > K: \\ \rho_{ij, k\ell} > 0}} D^{\delta}_{ij} - \frac{1}{\rho_{ij, k\ell}} D^{0}_{k \ell} \right) \right] -  \left[1 - \Phi\left(\min\limits_{\substack{k \leq K,\, \ell > K: \\ \rho_{ij, k\ell} < 0}} D^{\delta}_{ij} - \frac{1}{\rho_{ij, k\ell}} D^{0}_{k \ell} \right) \right]}. 
\end{equation}

Our proposed procedure of rejecting $\cup_{i \leq K, j > K}^n H^{\delta}_{ij}$ whenever \eqref{eq:rejection} holds corresponds exactly to rejecting when all these selective p-values are at most $\alpha$, establishing the validity of the procedure.   

\subsection{Proof of \Cref{thm:interpretation}}
\label{sec:interpretation_proof}

Let $I^{\delta} \in S$,  $J^{\delta} \in S$ be indices that minimize $D^{\delta}_{ij}$ over $i \in S$, $j \not \in S$, i.e., $D^{\delta}_{I^{\delta}J^{\delta}} = \min_{i \in S, j \not \in S} D^{\delta}_{ij}$. We will show that, defining $\delta^+ = \max(\delta, 0)$, the condition \eqref{eq:rejection} is satisfied whenever
\begin{equation}
    1 - \Phi(D^{\delta^+}_{I^{\delta^+}J^{\delta^+}}) \leq \alpha/2. 
\end{equation}

We again without loss of generality perform our analysis conditional on the specific event $S = \{1, \dots, K\}$. Again fix $i \leq K$ and $j > K$. Recall that if $a > b > c$, then $(b -c)/(a - c) \leq b/a$. Using this fact, we can bound every selective p-value \eqref{eq:selective_p_value}:

\begin{align*}
    &\frac{ \left[1 - \Phi(D^{\delta}_{ij} ) \right] - \left[1 - \Phi\left(\min\limits_{\substack{k \leq K,\, \ell > K: \\ \rho_{ij, k\ell} < 0}} D^{\delta}_{ij} - \frac{1}{\rho_{ij, k\ell}} D^{0}_{k \ell} \right) \right] }{\left[1 - \Phi\left(\max\limits_{\substack{k \leq K,\, \ell > K: \\ \rho_{ij, k\ell} > 0}} D^{\delta}_{ij} - \frac{1}{\rho_{ij, k\ell}} D^{0}_{k \ell} \right) \right] -  \left[1 - \Phi\left(\min\limits_{\substack{k \leq K,\, \ell > K: \\ \rho_{ij, k\ell} < 0}} D^{\delta}_{ij} - \frac{1}{\rho_{ij, k\ell}} D^{0}_{k \ell} \right) \right]}. \\
    &\leq \frac{ 1 - \Phi(D^{\delta}_{ij} )  }{1 - \Phi\left(\max\limits_{\substack{k \leq K,\, \ell > K: \\ \rho_{ij, k\ell} > 0}} D^{\delta}_{ij} - \frac{1}{\rho_{ij, k\ell}} D^{0}_{k \ell} \right) } \\
    &= \max\limits_{\substack{k \leq K,\, \ell > K: \\ \rho_{ij, k\ell} > 0}} \frac{ 1 - \Phi(D^{\delta}_{ij} )  }{1 - \Phi\left( D^{\delta}_{ij} - \frac{1}{\rho_{ij, k\ell}} D^{0}_{k \ell} \right) }. \\
\end{align*}

We restrict our attention to pairs $k \leq K$ and $\ell > K$  such that $\rho_{ij, k\ell} > 0$ and further bound the term inside the maximium in two separate cases.\newline 

\noindent \textbf{Case one:} $\boldsymbol{ D^{\delta}_{ij} \leq D^0_{k\ell}}$. Since $\frac{1}{\rho_{ij, k \ell}} \geq 1$, in this case we have $ D^{\delta}_{ij} - \frac{1}{\rho_{ij, k\ell}} D^0_{k \ell} \leq 0 $, so 
\begin{equation*}
    \frac{ 1 - \Phi(D_{ij}^{\delta})  }{1 - \Phi(D_{ij}^{\delta} - \frac{1}{\rho_{ij, k\ell}} D^0_{k \ell}) } \leq \frac{ 1 - \Phi(D^{\delta}_{ij})  }{1 - \Phi(0) } \leq 2(1 - \Phi(D^{\delta}_{I^{\delta}J^{\delta}}))
\end{equation*}

\noindent \textbf{Case two:} $\boldsymbol{ D^{\delta}_{ij} > D^0_{k\ell}}$. To handle this case we first show that 
\begin{equation*}
    f(x) = \frac{ 1 - \Phi(x)  }{1 - \Phi(x - \frac{1}{\rho_{ij, k\ell}} D^0_{k \ell}) }
\end{equation*}
is a non-increasing function of $x$. The derivative of the function is 
\begin{align*}
    f'(x) &= \frac{\phi(x) \phi\left(x - \frac{1}{\rho_{ij, j\ell}} D^0_{k \ell}\right)}{ \left(1 - \Phi\left(x - \frac{1}{\rho_{ij, k\ell}} D^{0}_{k \ell} \right)^2\right)} \left(\frac{1 - \Phi(x )}{\phi(x)} - \frac{1-\Phi\left(x - \frac{1}{\rho_{ij, k\ell}} D^0_{k\ell} \right)}{\phi\left(x - \frac{1}{\rho_{ij, k\ell}} D^0_{k\ell} \right)}  \right) \leq 0
\end{align*}
where the inequality follows from the fact that the Mills ratio $(1-\Phi(x))/\phi(x)$ is strictly decreasing \citep{Mills, Baricz}, and we always have 
\begin{equation*}
    x > x - \frac{1}{\rho_{ij, k\ell}} D^0_{k\ell}
\end{equation*}
because $\rho_{ij, k\ell} \geq 0$ and $D^0_{k\ell} \geq 0$. The non-positiveness of the derivative and the fact that $ D^{\delta}_{ij} > D^0_{k\ell}$ implies that 

\begin{align*}
    \frac{ 1 - \Phi(D_{ij}^{\delta})  }{1 - \Phi(D_{ij}^{\delta} - \frac{1}{\rho_{ij, k\ell}} D^0_{k \ell}) }  &\leq \frac{ 1 - \Phi(D^0_{k\ell})  }{1 - \Phi(D^0_{k\ell} - \frac{1}{\rho_{ij, k\ell}} D^0_{k\ell} ) } \\
    &\leq \frac{ 1 - \Phi(D^{0}_{k\ell})  }{1 - \Phi(0) } \\
    &\leq 2(1 - \Phi(D^{0}_{I^0J^0})). 
\end{align*}
where we have that $ D^0_{k\ell} - \frac{1}{\rho_{ij, k\ell}} D^0_{k\ell} \leq 0 $ because $\frac{1}{\rho_{ij, k \ell}} \geq 1$ and $D^0_{k\ell} \geq 0$. \newline 

If $\delta \geq 0$, we have that $D_{I^{\delta} J^{\delta}}^{\delta} \leq D^{\delta}_{k \ell} \leq D^{0}_{k \ell}$ for all $k \leq K$ and $\ell > K$. Thus $D^{\delta}_{I^{\delta}J^{\delta}} \leq D^{0}_{I^0 J^0}$ and we can combine our earlier two cases:
\begin{align*}
    &2(1 - \Phi(D^0_{I^0 J^0})) \leq 2(1 - \Phi(D^{\delta}_{I^{\delta} J^{\delta}}))\\
    &\implies \max\limits_{\substack{k \leq K,\, \ell > K: \\ \rho_{ij, k\ell} > 0}} \frac{ 1 - \Phi(D^{\delta}_{ij} ) }{1 - \Phi\left( D^{\delta}_{ij} - \frac{1}{\rho_{ij, k\ell}} D^0_{k \ell} \right)} \leq 2(1 - \Phi(D^{\delta}_{I^{\delta}J^{\delta}})). 
\end{align*} 
On the other hand, if $\delta < 0$, then $D^{0}_{I^0J^0} \leq D^{0}_{k\ell} \leq D^{\delta}_{k\ell} $ for all $k \leq K$ and $\ell > K$. Thus $D^{0}_{I^0J^0} \leq D^{\delta}_{I^{\delta} J^{\delta}}$ and we can again combine our earlier two cases: 
\begin{align*}
    &2(1 - \Phi(D^{\delta}_{I^{\delta} J^{\delta}})) \leq 2(1 - \Phi(D^{0}_{I^0J^0}))\\
    &\implies \max\limits_{\substack{k \leq K,\, \ell > K: \\ \rho_{ij, k\ell} > 0}} \frac{ 1 - \Phi(D^{\delta}_{ij} ) }{1 - \Phi\left( D^{\delta}_{ij} - \frac{1}{\rho_{ij, k\ell}} D^0_{k \ell} \right)} \leq 2(1 - \Phi(D^{0}_{I^0J^0})).
\end{align*} 
This is sufficient to imply the result.

\subsection{Proof of \Cref{cor:reduction}}
\label{sec:reduction_proof}

For this part of the argument, we fix $\delta=0$ and prove exactly the statement in \Cref{cor:reduction}. To show \Cref{cor:reduction}, we will show that, under the specified conditions, the left-hand side of \eqref{eq:rejection} is exactly equal to $2(1-\Phi(D_{IJ}))$, where $I$ and $J$ are from \Cref{thm:interpretation}'s statement and identical to $I^0$ and $J^0$ from the previous proof. 

If $\Sigma$ is diagonal or an equicorrelation matrix, then it is straightforward to check for $i, k \neq j, \ell$ that $\rho_{ij, k \ell} \geq 0$. This simplifies our procedure considerably. Again, without loss of generality, we perform our analysis conditional on $S = \{1, \dots, K$\}. Because, $\delta=0$ and $\rho_{ij, k \ell} \geq 0$ for $i, k\leq K$ and $j, \ell > K$, the condition \eqref{eq:rejection} is satisfied when
\begin{equation*}
    \max_{i \leq K, j > K} \frac{ 1 - \Phi(D_{ij}) }{1 - \Phi\left(\max\limits_{\substack{k \leq K,\, \ell > K: \\ \rho_{ij, k\ell} > 0}} D_{ij} - \frac{1}{\rho_{ij, k\ell}} D_{k \ell} \right)} \leq \alpha.
\end{equation*}
We still have the bound 
\begin{equation*}
    \max_{i \leq K, j > K} \frac{ 1 - \Phi(D_{ij}) }{1 - \Phi\left(\max\limits_{\substack{k \leq K,\, \ell > K: \\ \rho_{ij, k\ell} > 0}} D_{ij} - \frac{1}{\rho_{ij, k\ell}} D_{k \ell} \right)}  \leq 2(1-\Phi(D_{IJ})),
\end{equation*}
from proof of \Cref{thm:interpretation}. It is also the case that 
\begin{align*}
    \max_{i \leq K, j > K} \frac{ 1 - \Phi(D_{ij}) }{1 - \Phi\left(\max\limits_{\substack{k \leq K,\, \ell > K: \\ \rho_{ij, k\ell} > 0}} D_{ij} - \frac{1}{\rho_{ij, k\ell}} D_{k \ell} \right)} &\geq \frac{ 1 - \Phi(D_{IJ} ) }{1 - \Phi\left(\max\limits_{\substack{k \leq K,\, \ell > K: \\ \rho_{ij, k\ell} > 0}} D_{IJ} - \frac{1}{\rho_{IJ, k\ell}} D_{k \ell} \right)}\\
    &\geq  \frac{ 1 - \Phi(D_{IJ}) }{1 - \Phi( D_{IJ} - \frac{1}{\rho_{IJ, IJ}} D_{IJ})}\\
    &=  2(1 - \Phi(D_{IJ})),
\end{align*}
so the two expressions are in fact equal. Thus, when $\delta=0$ and we are in the specified settings, the procedure from \Cref{thm:method} rejects exactly when 
\begin{equation*}
    1 - \Phi(D_{IJ}) \leq \alpha/2, 
\end{equation*}
as desired.

\section*{Acknowledgements}
The author would like to thank John Cherian, Tim Morrison, Yash Nair, and James Yang for helpful discussions. 

\bibliographystyle{plainnat}
\bibliography{bibliography.bib}

\begin{appendix}

\section{Losing power for certain covariance structures}
\label{sec:counterexample_appdx}

\Cref{thm:method}'s procedure can be more powerful than just checking the condition in \Cref{thm:interpretation} when for $i \neq j$ and $k \neq \ell$, correlations between pairs of differences $X_i - X_j$ and $X_k - X_{\ell}$ are negative. We will consider the $K=1$ problem, and generate data

\begin{align*}
    &X_1 = \sigma_1 Z_1 + \mu_1, \\
    &X_2 = \sigma_2 Z_2 + \mu_2, \\ 
    &X_j = -\sigma_2 Z_2 + \sigma_3 Z_j + \mu_j \text{ for } j > 2,
\end{align*}
where $Z_i$ are independent standard Gaussian random variables.

If we set $\mu_1 > \mu_2 > \mu_3 = \dots = \mu_n$ properly,  then the top $K = 1$ set $S$ will often be $\{1\}$, and the maximum in \Cref{thm:method}'s condition will often be achieved by $i=1, j = 2$. By setting $\sigma_2 > \sigma_1$ and $\sigma_2 > \sigma_3$, we can ensure that $\rho_{12, 1j}$ is very negative for $j > 2$. This will mean there is a lot of benefit to running \Cref{thm:method}'s full test in place of \Cref{thm:interpretation}'s simpler approach. We instantiate this problem by setting $n=5$, $\sigma_1^2 = 1$, $ \sigma^2_2 = 5$, $\sigma_3^2 = 0.1$, $\mu_1 = 5$, $\mu_2 = 3$, $\mu_3 = \mu_4 = \mu_5 = 0$. Over $B=10000$ simulated trials run at level $\alpha=0.1$, \Cref{thm:method}'s test has an empirical power of $1.0$, whereas the empirical power of \Cref{thm:interpretation}'s simpler approach is just below $0.057$. Here, power is defined to be the probability of rejecting conditional on $S = \{1\}$. 

\section{Some special cases}
\label{sec:special_appdx}

We document some special cases where we can better characterize the behavior of our method. 

\subsection{Equicorrelation}

Fix $\delta \geq 0$. We will argue that if $\Sigma$ is an equicorrelation matrix, i.e., $\Sigma_{ii} = \sigma^2$ and $\Sigma_{ij} = \rho \sigma^2$ when $i \neq j$, then $I^{\delta}$ and $J^{\delta}$ from the proof of \Cref{thm:interpretation} are the indices of the $K$ and $(K+1)$ largest entries of $X$ respectively. 

The indices $I^{\delta}$ and $J^{\delta}$ minimize 
\begin{equation*}
    D^{\delta}_{ij} = \frac{(X_i  - X_j) - \delta }{\sqrt{2\sigma^2 - 2\rho\sigma^2} }
\end{equation*}
over $i \in S$ and $j \not \in S$. This is clearly minimized if we take $i$ to be the index of the smallest entry in $S$ (i.e., the $K$th largest entry) and $j$ to be the index of the largest entry outside of $S$ (i.e., the $(K+1)$st largest entry).

\subsection{Autocorrelation}

We will show if $\Sigma$ is an autocorrelation matrix for an $\text{AR}(1)$ process, i.e., $\Sigma_{ij} = \sigma^2 \rho^{|i - j|}$ for some $\rho \in (-1, 1)$, then the conclusion of \Cref{cor:reduction} still holds so long as $K=1$ or $K = n -1$.

We show the result for $K=1$. The argument for $K= n-1$ is analogous.  When $K=1$ the proof of \Cref{cor:reduction} would go through so long as we had for $i \neq j, \ell$ that $\rho_{ij, i\ell} \geq 0$ . It suffices to ensure that

\begin{align*}
    c_{ij, i\ell} &= \Cov(X_i - X_j, X_i - X_{\ell}) \\
                  &= \Sigma_{ii} - \Sigma_{ij} - \Sigma_{i\ell} + \Sigma_{j\ell}\\
                  &= \sigma^2  - \sigma^2 \rho^{|i- j|} - \sigma^2 \rho^{|i- \ell|} + \sigma^2 \rho^{|j - \ell|}\\
\end{align*}
is at least zero. When $\rho \geq 0$ we have that $c_{ij, i\ell} \geq \sigma^2(1 - 2\rho)$. When $\rho \leq 0$, we have that $c_{ij, i\ell} \geq \sigma^2(1 - |\rho| - \rho^2 )$. Thus, as long as $\rho \in [\frac{1-\sqrt{5}}{2}, \frac{1}{2}]$, we have $c_{ij, i\ell} \geq 0$. This is achieved whenever $|\rho| \leq 1/2$. 

\subsection{Multinomial}

Let $Y \sim \text{Multinomial}(t, \pi_1, \dots, \pi_n)$ and suppose we define $\hat{\pi} = Y/t$ and use a Gaussian approximation $\hat{\pi} \sim N(\pi, \hat{\pi}/t - \hat{\pi}\hat{\pi}^{\top}/t)$. Consider applying our method to the observation $\hat{\pi}$ while using the covariance $\Sigma = \hat{\pi}/t - \hat{\pi}\hat{\pi}^{\top}/t$.  We will show that the conclusion of \Cref{cor:reduction} still holds so long as $K=1$.

Without loss of generality, suppose that $\hat{\pi}$ is in sorted order, so $\hat{\pi}_1 \geq \dots \geq \hat{\pi}_n$. The proof of \Cref{cor:reduction} would go through so long as we had for $j, \ell > 1$ that $\rho_{1j, 1\ell} \geq 0$ . It suffices to see then that 

\begin{align*}
    c_{ij, i\ell} &= \Cov(X_i - X_j, X_i - X_{\ell}) \\
                  &= \Sigma_{ii} - \Sigma_{ij} - \Sigma_{i\ell} + \Sigma_{j\ell}\\
                  &= \frac{1}{t}(\hat{\pi}_{1}(1-\hat{\pi}_1) + \hat{\pi}_1\hat{\pi}_j + \hat{\pi}_{1}\hat{\pi}_{\ell}  - \hat{\pi}_j \hat{\pi}_{\ell})\\
                  &= \frac{1}{t}(\hat{\pi}_{1}(1-\hat{\pi}_1) + \hat{\pi}_1\hat{\pi}_j + (\hat{\pi}_{1} - \hat{\pi}_j)\hat{\pi}_{\ell}) \\
                  &\geq 0
\end{align*}

Fixing $\delta \geq 0$, we show in this same setting that, when $K=1$, the indices $I^{\delta}$ and $J^{\delta}$ from the proof of \Cref{thm:interpretation} are always the indices of the largest and second largest entries of $\hat{\pi}$ respectively. Because $K=1$ we know that $I^{\delta} = 1$ will be the index of the largest entry. The index $J^{\delta}$ must then minimize
\begin{equation*}
    D^{\delta}_{1j} = \frac{\sqrt{t}[(\hat{\pi}_1 - \hat{\pi}_j) - \delta]}{\sqrt{\hat{\pi}_1(1- \hat{\pi}_1) + \hat{\pi}_j(1- \hat{\pi}_j) + 2\hat{\pi}_1 \hat{\pi}_j}} = \frac{\sqrt{t}[(\hat{\pi}_1 - \hat{\pi}_j) - \delta]}{\sqrt{\hat{\pi}_1 + \hat{\pi}_j - (\hat{\pi}_1 - \hat{\pi}_j)^2}}
\end{equation*}
over $j > 1$. It is clear that setting $J^{\delta} = 2$ will minimize the numerator of $D^{\delta}_{1j}$ and also maximize its denominator, which suffices to establish our claim. 

\section{Simultaneous approach}
\label{sec:simul_appdx}

For the sake of comparison, we derive a simultaneous inference approach for the problem of drawing the inference $ \min_{i\in S}\mu_i > \max_{j \not \in S} \mu_j $. We base our approach off of that in \cite{Bofinger1983, Bofinger1985} and \cite{Hsu1981}, which assume an isotropic covariance and derive an approach that dominates Tukey's HSD test. To handle the case with general covariances, we will have to come up with a slightly different procedure than what currently exists in the literature. It is not tractable, but it still helps inform us of the limits of simultaneous inference. 

Let $Z = X -\mu$ be a centered version of $X$, so that $Z \sim N(0, \Sigma)$, and $\Pi$ denote the set of permutations over $n$ elements. Defining 
\begin{equation}
    \label{eq:quantile}
    q_{1-\alpha} = \max_{\pi \in \Pi} \text{Quantile}\left(1-\alpha, \max\left\{\frac{Z_{\pi^{-1}(K)} - Z_{\pi^{-1}(K+1)}}{v_{\pi^{-1}(K), \pi^{-1}(K+1)}}, \max\limits_{\substack{ i \geq K + 1, \\ j \leq K}} \frac{Z_{\pi^{-1}(i)} - Z_{\pi^{-1}(j)}}{v_{\pi^{-1}(i), \pi^{-1}(j)}}  \right\}    \right),
\end{equation}
we argue that we can safely draw the inference $ \min_{i\in S}\mu_i > \max_{j \not \in S} \mu_j $ whenever 
\begin{equation*}
    X_I \geq X_J + v_{IJ} q_{1-\alpha}, 
\end{equation*}
where $I$ and $J$ are as defined in the statement of \Cref{thm:interpretation}. 

Define $I'$ to be the largest index in $S$ (corresponding to the smallest mean) and $J'$ to be the smallest index not in $S$ (corresponding to the largest mean). A false rejection happens exactly when $X_I > X_J + v_{IJ} q_{1-\alpha}$ and also $\mu_{I'} - \mu_{J'} \leq 0$. Note that $I' \geq K$, and $I' = K \implies J' = K + 1$ and $I' > K \implies J' \leq K$. With this in mind, we can bound  

\begin{align*}
    P(\text{false rejection}) &= P\left(\mu_{I'} - \mu_{J'} \leq 0, \frac{X_I - X_J}{v_{IJ}} - q_{1-\alpha}\geq 0 \right)\\
    &\leq P\left( \frac{\mu_{I'} - \mu_{J'}}{v_{I'J'}} \leq 0, \frac{X_{I'} - X_{J'}}{v_{I'J'}} - q_{1-\alpha}\geq 0 \right)\\
    &\leq P\left( \frac{\mu_{I'} - \mu_{J'}}{v_{I'J'}} \leq \frac{X_{I'} - X_{J'}}{v_{I'J'}} - q_{1-\alpha} \right)\\
    &\leq P\left(q_{1-\alpha} \leq \frac{X_{I'} - \mu_{I'}}{v_{I'J'}} - \frac{X_{J'} - \mu_{J'} }{v_{I'J'}}\right) \\
    &\leq P\left(q_{1-\alpha} \leq \frac{Z_{I'} - Z_{J'}}{v_{I'J'}}\right) \\
    &\leq P\left(q_{1-\alpha} \leq \max\left\{\frac{Z_K - Z_{K+1}}{v_{K, K+1}}, \max\limits_{\substack{i \geq K + 1, \\ j \leq K}} \frac{Z_i - Z_j}{v_{ij}}  \right\}  \right) \\
    &\leq \alpha 
\end{align*}
where the last inequality follows from the definition of $q_{1-\alpha}$.

Having proven the validity of a simultaneous method, we make two points. First, we could not use the $1-\alpha$ quantile of 
    \begin{equation*}
        \max\left\{\frac{Z_K - Z_{K+1}}{v_{K, K+1}}, \max\limits_{\substack{i \geq K + 1, \\ j \leq K}} \frac{Z_i - Z_j}{v_{ij}}  \right\}
    \end{equation*}
 in our procedure because, in the non-isotropic covariance case, computing this quantile requires us to know how to order the samples by their means. Second, $q_{1-\alpha}$ is certainly at most 
    \begin{equation}
        \label{eq:hsd_quantile}
         h_{1-\alpha} = \text{Quantile}\left(1-\alpha,  \max_{i \neq j} \frac{|Z_i - Z_j|}{v_{ij}}  \right),
    \end{equation}
which justifies the validity of the Tukey's HSD variant we proposed in the main text.

Now, let's compare \Cref{thm:method}'s test to this simultaneous approach. It is easy to see that $q_{1-\alpha} \geq z_{1-\alpha/2}$, 
where the inequality is strict whenever $n > 2$. Thus \Cref{thm:method}'s test has a rejection region that matches the simultaneous approach's rejection region when $n = 2$, and is a strict superset of it when $n > 2$. 

In the case that the $X_i$ are independent, we can more explicitly quantify the difference in the two rejection regions. Suppose that $n$ is arbitrarily large and, without loss of generality, that $K < n/2$ (the case that $K \geq n/2$ can be handled with an identical argument). Let $\pi$ be the permutation such that $\pi^{-1}(K)$ satisfies $\Sigma_{\pi^{-1}(K), \pi^{-1}(K)} \leq \Sigma_{m, m}$ for all $m$. Then, 

\begin{align*}
    &\max\left\{\frac{Z_{\pi^{-1}(K)} - Z_{\pi^{-1}(K+1)}}{v_{\pi^{-1}(K), \pi^{-1}(K+1)}}, \max\limits_{\substack{ i \geq K + 1, \\ j \leq K}} \frac{Z_{\pi^{-1}(i)} - Z_{\pi^{-1}(j)}}{v_{\pi^{-1}(i), \pi^{-1}(j)}}  \right\} \\
    &\geq \max_{i \geq K + 1} \frac{Z_{\pi^{-1}(i)} - Z_{\pi^{-1}(K)} }{v_{\pi^{-1}(i), \pi^{-1}(K)}} \\
    &\geq \max_{i \geq K + 1} \frac{Z_{\pi^{-1}(i)}}{v_{\pi^{-1}(i), \pi^{-1}(K)}} - \max_{i \geq K + 1} \frac{Z_{\pi^{-1}(K) }}{v_{\pi^{-1}(i), \pi^{-1}(K)}} \\
    &\geq \max_{i \geq K + 1} \frac{Z_{\pi^{-1}(i)}}{v_{\pi^{-1}(i), \pi^{-1}(K)}} - \max_{i \geq K + 1} \frac{Z_{\pi^{-1}(K)} I(Z_{\pi^{-1}(K)} > 0) }{v_{\pi^{-1}(i), \pi^{-1}(K)}} \\
    &= \max_{i \geq K + 1} \frac{Z_{\pi^{-1}(i)} I(Z_{\pi^{-1}(i)} > 0) }{v_{\pi^{-1}(i), \pi^{-1}(K)}} -  \frac{Z_{\pi^{-1}(K)} I(Z_{\pi^{-1}(K)} > 0) }{\sqrt{2} \sqrt{\Sigma_{\pi^{-1}(K), \pi^{-1}(K)}} } + o_p(1) \\
    &\geq \max_{i \geq K + 1} \frac{Z_{\pi^{-1}(i)} I(Z_{\pi^{-1}(i)} > 0) }{\sqrt{2} \sqrt{\Sigma_{\pi^{-1}(i), \pi^{-1}(i) }}} + O_p(1) \\
    &\geq \frac{1}{\sqrt{2}} \max_{i \geq K + 1} \frac{Z_{\pi^{-1}(i)}}{ \sqrt{\Sigma_{\pi^{-1}(i), \pi^{-1}(i) }}} + O_p(1) \\
    &= O(\sqrt{\log n}) + O_p(1)\\
\end{align*}
where the last equality follows from applying standard extreme value theory results regarding the concentration of the maximum of independent standard Gaussians \citep{Haan} and the fact that $n - K \geq n/2$ per our assumption. As a consequence, $q_{1-\alpha}$ must grow at least on the order of $\sqrt{\log n}$ as well. This implies that, in the independent case, the HSD quantile \eqref{eq:hsd_quantile} grows at least on the order of $\sqrt{\log n}$ also. 

\section{Getting a confidence lower bound}
\label{sec:clb_appdx} 

By inverting the test \eqref{eq:rejection} for different values of $\delta$ (i.e., considering the set of $\delta$ for which we fail to reject), we get a confidence region for the gap $\min_{i \in S} \mu_i - \max_{j \neq S} \mu_j $ between the smallest mean in the selected set and the largest mean in the unselected set that is valid conditional on $S$. It is not immediately clear, however, that this region will result in a confidence lower bound (i.e., there is some smallest $\delta$ for which we fail to reject). We provide an argument that it does. 

Appendix B.3 of \cite{Sood} tells us that, because our original marginal p-values $p^{\delta}_{ij}$ in \eqref{eq:p_value} come from the UMP test in a MLR family and because our selection event does not depend on the parameter $\delta$ we are testing, the selective p-values $p^{\delta}_{sel, ij}$ from \eqref{eq:selective_p_value} are non-decreasing in $\delta$. If, for $i \in S$ and $j \not \in S$ we define 

\begin{equation*}
\hat{\mu}_{ij} =
\begin{cases}
    \infty , & p^{\delta}_{sel, ij} < \alpha \text{ for all } \delta, \\
    -\infty, & p^{\delta}_{sel, ij} > \alpha \text{ for all } \delta, \\
    \sup \{\delta: p^{\delta}_{sel, ij} = \alpha \} & \text{otherwise}, 
\end{cases}
\end{equation*}
then it is straightforward to argue that \Cref{thm:method}'s procedure will fail to reject if and only if $\delta > \min_{i \in S, j \not \in S} \hat{\mu}_{ij}$. Therefore the inverted confidence region does indeed correspond to a confidence lower bound. 

Recalling $I^{\delta}$ and $J^{\delta}$ from the proof of \Cref{thm:interpretation}, the more general result we prove in \Cref{thm:interpretation} implies that the following more computationally easier confidence lower bound for $\min_{i \in S} \mu_i - \max_{j \neq S} \mu_j $ is still valid conditional on $S$.
First, check if $1- \Phi(D^{0}_{I^0 J^0} ) > \alpha/2$. If so, return $-\infty$. Otherwise, return the $\delta$ for which $\alpha/2 = 1 - \Phi(D^{\delta}_{I^{\delta}J^{\delta}})$. Noting that 
\begin{equation*}
   \Phi(D^{\delta}_{I^{\delta}J^{\delta}}) = \min_{i \in S, j \not \in S} 1 - \Phi\left( \frac{X_i - X_j - \delta}{ v_{ij}}\right)
\end{equation*}
is the minimum of a finite number of  Gaussian p-values from UMP one-sided testing, there will be some first time that this minimum equals $\alpha/2$. Keep in mind that, while easier to compute, this confidence lower bound will always be at least as large as the one that results from inverting the test \eqref{eq:rejection} (this is an implication of the proof of \Cref{thm:interpretation}). 

\end{appendix}

\end{document}